\documentclass{article}

\usepackage{amssymb,amsfonts,amsmath}
\usepackage{cite,enumerate,float}
\usepackage{color}
\usepackage{tikz}
\usetikzlibrary{arrows,snakes,backgrounds}

\usepackage[utf8]{inputenc}
\usepackage[russian]{babel}

\def\be{\begin{eqnarray}}
\def\ee{\end{eqnarray}}
\def\nn{\nonumber}

\def\p{\partial}
\def\tr{{\rm tr}\,}
\def\Tr{{\rm tr}\,}

\definecolor{red}{rgb}{1,0,0}
\definecolor{orange}{rgb}{1,0.5,0}
\definecolor{violet}{rgb}{0.7,0,1}

\hoffset= - 1.0in         

\begin{document}

\hfill MIPT/TH-05/26

\hfill IITP/TH-05/26

\hfill ITEP/TH-05/26

\bigskip

\bigskip

\centerline{\Large{
Averages of Exponentials from the point of view of Superintegrability
}}

\bigskip

\bigskip

\centerline{\bf A.Morozov}

\bigskip

\centerline{ {\it MIPT, Dolgoprudny, 141701, Russia}}
\centerline{ {\it NRC ``Kurchatov Institute", 123182, Moscow, Russia}}
\centerline{ {\it Institute for Information Transmission Problems, Moscow 127994, Russia}}
\centerline{ {\it ITEP, 117259, Moscow, Russia}}

\bigskip

\bigskip

\centerline{ABSTRACT}

\bigskip

{\footnotesize
We calculate Gaussian averages of arbitrary exponentials of the matrix variable $X$
with the help of superintegrability, which provides explicit expressions for Schur averages.
As in the simpler cases the answer is expressed in terms of Laguerre polynomials,
but in a somewhat sophisticated way.
It involves triangular sum over partitions, with simple exponential factor
and a complicated polynomial prefactor.
Some ingredients of the formula are not found in full generality and there is still a room
for further work.
}

\bigskip

\bigskip

\section{Introduction}

Eigenvalue matrix models remain in the center of studies in theoretical physics
since they first appeared in \cite{Wigner,Dyson,Mehta}.
As time goes, we understand that they capture the most important features of
string theory -- and thus of phenomenologically important quantum field theory.
And they separate the complexity of these theories from the basic properties,
which are truly important -- perhaps, not for applications, but for understanding.
This is especially important in the non-perturbative domain, where matrix models
-- along with their close relatives, the supersymmetric models --
still remain the main sources of precise and reliable knowledge.
In particular they stand behind the discovery \cite{GKMMM,UFN23,Mir}
of the central role of integrability in non-perturbative quantum physics
(which just reflects the possibility to change integration variables on functional integrals),
and probably point out to an even stronger property --
the superintegrability \cite{SI}, which implies the existence of variables,
where integrals can be taken exactly, {\it a la} Duistermaat-Heckman theorem \cite{DH} or
``exact quasiclassics'' \cite{Niemi}.
It is amusing how rich are even the simplest eigenvalue models,
and how new information is being constantly pumped from them.

One of the old questions about the simplest Gaussian model with the action $e^{-\tr X^2}$
concerns calculating the averages of exponentials ${\rm Tr}_R\, e^{\lambda {\cal X}}$
in various representations $R$.
While this question for polynomials $\pi_k=\tr X^k$ is resolved by superintegrability --
the proper basis is provided by Schur polynomials of $\pi_k$, -- for exponentials it remains open.
In the classical paper  \cite{0010274} the simplest such average, of just $\tr e^{\lambda X}$
in the fundamental representation,
was evaluated by the artful application of orthogonal-polynomial technique \cite{Mehta,orthopol} --
and it turned out to be a Laguerre polynomial.
Later this calculation was extended to symmetric and antisymmetric averages in \cite{1808.10161},
and the answers are quite complicated.
It is, perhaps, time to find a general average in any representation -- and this is the question
addressed in the present paper.
We approach this problem by the modern technique of superintegrability,
and sec.2 below can serve as an instruction of how it is applied to practical calculations --
in fact, not restricted to the exponential case.
We do not find a full answer for the exponential average,
but deduce its peculiar structure in the form of triangular decomposition (\ref{sigmadeco})
and explain the properties of its particular ingredients.

\bigskip

If these general theoretical questions do not look stimulative enough,
it deserves mentioning some more concrete applications -- of which we choose two.

First, in many senses the exponential averages are associated with Wilson loops in Yang-Mills theories.
In the framework of this analogy one can wonder about the physical significance of Wilson-loops
in non-trivial representations.
This parameter -- representation -- is often ignored in the standard considerations,
because quarks in QCD are in the fundamental one, and why should one be interested in anything else?
However, even at this level of consideration, representation matters.
If we consider multi-quark systems -- what we probably need to do in the study, say, of pentaquarks,
then one of the options is that the pairs and triples are not themselves singlets, but belong
to less trivial representation of $sl_3$ \cite{MMpenta}.
In fact, the representation-dependence of Wilson loop averages in Yang-Mills theories is
quite pronounced -- it is enough to say that confinement disappears in the adjoint sector.
At a much deeper level this poses a question of relation between confinement and Vogel theory
\cite{Vogel,MSle}, which also distinguishes adjoint sector among the set of all representations.

Second, the problem of exponential averages in Gaussian matrix model is rather popular,
because holography relates them to branes in ${\rm S}^5\times{\rm AdS}_5$ space
\cite{Mal,ESZ,0010274}.
Exact matrix-model answers should be compared with string corrections to the classical brane area
in Polyakov formalism, see \cite{BKTs,MiZa} for the latest advances and previous references.
At both sides one can consider arbitrary representations and the full duality requires matching
between all of them.
We do not touch this relation in the present paper,
just note that appearance of Laguerre calculus on the string side would be quite an interesting
result -- still to be revealed.

\bigskip

We begin in sec.\ref{si} from derivation of exponential correlators $\sigma_R$ from the polynomial ones,
dictated by superintegrability formulas.
This is a technically important piece, which can be also useful for other kinds of non-perturbative
calculations.
However, it provides a rather sophisticated expression (\ref{sigmafromsi}),
which for our goals is just a source of formulas for further analysis.
We use it to generate numerous examples, the simplest of them are presented in sec.\ref{exa},
some more are in the attachment {\it expoave.txt} to this submission.
Looking through these examples, one can reveal the structure of the answer for $\sigma_R$ in the form
of triangular decomposition (\ref{sigmadeco}) -- which is made out of Kostka matrices, certain eigenvalues
and peculiar polynomials $P_Q$.
This decomposition is described in the sec.\ref{exasta}.
The polynomials are described in a separate sec,\ref{pols}, where they are expressed through
Laguerre matrix ${\cal A}$ (\ref{Amatrix}), introduced in \cite{1808.10161} and further studied in recent \cite{MOk}.
In fact this description directly provides the products $\mathfrak{A}_Q$ of eigenvalues and $P_Q$'s,
what allows further simplification of  (\ref{sigmadeco}) to (\ref{sigmadecoU})
and leads to the final formulation of our results in (\ref{summary}).
In between (\ref{sigmadeco}) and (\ref{sigmadecoU}) be insert reformulation in these terms
of our original examples from sec.\ref{exa} -- this can be useful for a deeper insight into the
formulas, but can also be omitted as a superficial comment.
We end with enumerating some drawbacks of our answer in the conclusion sec.\ref{conc} --
with the hope that they can be cured by the further studies in this area.
Applications are not touched, as we already mentioned they can be diverse
and constitute other research directions.

\section{Schur of exponential from superintegrable Schur average
\label{si}}

The Gaussian integral over $N\times N$ Hermitian matrices $X$,
\be
\langle F \rangle \ \sim \int F\{\tr X^k\} e^{-\frac{1}{2} \tr X^2}  dX
\label{Gamodel}
\ee
for $F\{p_k=\tr X^k\}$, normalized so that $\langle 1\rangle =1$,
possesses the ``superintegrability'' property \cite{SI}
\be
\langle S_R \rangle = \eta_R S_R[N]
\label{aveSchur}
\ee
for Schur functions   $S_R\{p\}$ with $S_R[N] := S_R\{p_k=N\}$
and
\be
\eta_R := \frac{S_R\{p_k=\delta_{k,2}\}}{S_R\{p_k=\delta_{k,1}\}}
\ee

The question to be addressed in this paper is to evaluate
\be
\boxed{
\sigma_Q := \left<{\rm Tr}_Q e^{\lambda {\cal X}}\right> = \left< S_Q\!\left[e^{\lambda X}\right]\right>
}
\ee
Here
$S_Q\!\left[e^{\lambda X}\right]:=S_Q\{p_k = \tr e^{k\lambda X} \}$.

\bigskip

If we wish to use (\ref{aveSchur}) as a definition of Gaussian average,
then we need to represent arbitrary $F\{p_k\}$ in (\ref{Gamodel}) as a linear combination
of Schur functions:
\be
F\{p_k\} = \sum_R S_R\{p_k\}\ \cdot\, \vdots\hat S_R F\vdots
\label{Cauchy}
\ee
where the dual Schur operators are $\hat S_R := S_R\left\{k\frac{\p}{\p p_k}\right\}$
and triple dots (``normal ordering'') in $\vdots\hat S_R F\vdots$ means
that we put all $p_k=0$ after that action
of derivatives, i.e. they act ``in full'', e.g.  $\vdots\frac{\p}{\p p_1} p_1^m\vdots = m \delta_{m,1}$.
Identity (\ref{Cauchy}) is a direct corollary of Cauchy formula for Schur functions,
\be
\sum_R S_R\{p\}S_R\{\bar p\} = \exp\left(\sum_k \frac{p_k\bar p_k}{k}\right)
\ee
From (\ref{Cauchy}) and (\ref{aveSchur})
we obtain:
\be
\langle F\rangle = \sum_R   \langle S_R\rangle \ \vdots\hat S_R F\vdots
= \sum_R \eta_R S_R[N] \ \vdots\hat S_R F\vdots
\ee
The next step is to calculate $\vdots\hat S_R S_Q\!\left[e^{\lambda X}\right]\vdots$
for our particular choice   $F=S_Q\!\left[e^{\lambda X}\right]$.

First  of all,
\be
S_Q\!\left[e^{\lambda X}\right]:=S_Q\{p_k = \tr e^{k\lambda X} \}
= S_Q\left\{ p_k= N + \sum_{n=1}^\infty  \frac{(k\lambda)^n}{n!}\tr X^n\right\}
\label{SQpi}
\ee
is a highly non-linear function of $\pi_n:=\tr X^n$.
We use this notation $\pi_n$ to avoid confusion between $p_k$ and $\tr X^k$.

To handle the calculation further we can use the expansion
\be
S_R\{p_k\} = \sum_{\Delta:\ |\Delta|=|R|} \frac{\chi(R,\Delta)}{z_\Delta} p_\Delta
\ee
with symmetric-group characters $\chi(R,\Delta)$
and the standard combinatorial factor
$z_\Delta:=\prod_k k^{\nu_k} \nu_k!$
with $\nu_k$ extracted from $p_\Delta = \prod_{i=1}^{l_\Delta} p_{\Delta_i}=\prod_k p_k^{\nu_k}$
for the partition (Young diagram)
$\Delta = [\Delta_1\geq \ldots \geq \Delta_{l_\Delta} > 0]$.
We need it in the form
\be
\vdots \hat S_R F \vdots =  \sum_{ |\Delta|=|R|} \frac{\chi(R,\Delta)}{z_\Delta}\
\vdots \left(\prod_{i=1}^{l_\Delta} \Delta_i \frac{\p}{\p p_{\Delta_i}}\right) F\vdots
\ee
with  $F$ from (\ref{SQpi}),
i.e.
\be
\vdots \hat S_R S_Q\left[e^{\lambda X}\right] \vdots =
\sum_{\stackrel{\Delta}{ |\Delta|=|R|}} \frac{\chi(R,\Delta)}{z_\Delta}\
\vdots \left(\prod_{i=1}^{l_\Delta} \Delta_i \frac{\p}{\p \pi_{\Delta_i}}\right)
S_Q\left\{ N + \sum_{n=1}^\infty  \frac{(k\lambda)^n}{n!}\pi_n\right\} \vdots
\ee

Finally
\be
\boxed{
\sigma_Q:= \left< S_Q\!\left[e^{\lambda X}\right]\right>
= \sum_R \eta_R S_R[N]
\sum_{\stackrel{\Delta}{ |\Delta|=|R|}} \frac{\chi(R,\Delta)}{z_\Delta}\
\vdots \left(\prod_{i=1}^{l_\Delta} \Delta_i \frac{\p}{\p \pi_{\Delta_i}}\right)
S_Q\left\{ N + \sum_{n=1}^\infty  \frac{(k\lambda)^n}{n!}\pi_n\right\} \vdots
}
\label{sigmafromsi}
\ee
This  formula can look complicated, still it is easily programmable.
The file expoave.txt, attached to  this submission, contains $\langle\sigma_Q\rangle$ in the form
of $N$-dependent series in $\lambda$ up to the order $O(\lambda^{20}$) for $|Q|\leq 6$.

\section{Mnemonic (computer-experiment) results
\label{exa}}

For $Q=[1^m]$ we obtain
\be
\sigma_{[1^m]} = e^{\frac{m}{2}\lambda^2} \cdot P_{[1]^m}(\lambda^2)
\label{aspolred}
\ee
but for $Q=[m]$ there is no such simple polynomial reduction.

For $m=1$ expression for (\ref{aspolred}) was deduced by direct orthogonal-polynomial calculation in
\cite{0010274}, and in this case $P_{N-1}$ is actually the Laguerre polynomial.

In more detail,
\be
\sigma_{[1]} =  e^{\frac{\lambda^2}{2}}L^1_{N-1}(-\lambda^2), \nn \\
\sigma_{[1,1]} =  e^{\lambda^2}P^{[1,1]}_N( \lambda^2), \nn \\
\sigma_{[2]} =  e^{2\lambda^2 }L_{N-1}(1,-4\lambda^2) + e^{\lambda^2}P^{[1,1]}_{N}(\lambda^2), \nn \\
\ldots
\label{simplestave}
\ee
with
\be
L^1_{0 }(-\lambda^2) = 1,\nn \\
L^1_{1}(-\lambda^2) = 2+\lambda^2, \nn \\
L^1_{2}(-\lambda^2) = 3+3\lambda^2+\frac{1}{2}\lambda^4, \nn \\
L^1_{3}(-\lambda^2) = 4 +6\lambda^2+2\lambda^4+\frac{1}{6}\lambda^6, \nn \\
L^1_{4}(-\lambda^2) = 5 +10\lambda^2+5\lambda^4+\frac{5}{6}\lambda^6+\frac{1}{24}\lambda^8, \nn \\
L^1_{5}(-\lambda^2) = 6 +15\lambda^2+10\lambda^4+\frac{5}{2}\lambda^6+\frac{1}{4}\lambda^8+\frac{1}{120}\lambda^{10},
 \nn \\
\ldots
\ee
and
\be
P^{[1,1]}_1 = 0,\nn \\
P^{[1,1]}_2 = 1, \nn \\
P^{[1,1]}_3 = 3+3\lambda^2+\frac{1}{2}\lambda^4, \nn \\
P^{[1,1]}_4 = 6 + 12\lambda^2+7\lambda^4+\frac{4}{3}\lambda^6+\frac{1}{12}\lambda^8, \nn \\
P^{[1,1]}_5 = 10 + 30\lambda^2+30\lambda^4+\frac{25}{2}\lambda^6+\frac{19}{8}\lambda^8
+\frac{5}{24}\lambda^{10}+\frac{1}{144}\lambda^{12}, \nn \\
\ldots
\ee
$L_N^a$ satisfy 3-term relations and are orthogonal polynomials (actually -- with the measure $x^ae^{-x}dx$).
But $P_N^{[1,1]}$ do {\it not} satisfy 3-term  relations are not {\it just} orthogonal polynomials.
It is our task to find what they are --
and what are their further generalizations for more complicated representations.

$P^{[1,1]}_N$ drops away from the combination
\be
\langle \pi_2\rangle = \sigma_{[2]}-\sigma_{[1,1]}
= e^{2\lambda^2} L_{N-1}^1(-4\lambda^2)
\label{pi2}
\ee
Generalization is the single-hook sum rule:
\be
\langle\pi_m\rangle =\left<\tr e^{m\lambda X}\right> =  \sum_{i=0}^{m-1} (-)^i\sigma_{[m-1,1^i]} =
e^{\frac{m^2}{2}\lambda^2} L_{N-1}^1(-m^2\lambda^2)
\label{pim}
\ee
This is an obvious corollary of the first formula in (\ref{simplestave}),
because the $\langle\pi_m\rangle =\left<\tr e^{m\lambda X}\right>$
is obtained by the substitution of $m\lambda$ for $\lambda$ in $\langle \pi_i \rangle$.

To simplify notations below we suppress $N$ dependence and denote
$P^{[1,1]}_N(\lambda^2)=P_{[1,1]}$,
so that
\be
\sigma_{[2]} = L_{N-1}^1(-4\lambda^2) e^{2\lambda^2 } + P_{[1,1]}e^{\lambda^2}
\ee
Also, $L_{N-1}^1(-\lambda^2)=P_{[1]}$.
This can be generalized:
\be
\sigma_{[1^m]} = P_{[1^m]}\cdot e^{\frac{m}{2}\lambda^2}
\ee
and
\be
\sigma_{[1,2]} - 2\sigma_{[1,1,1]} = P_{[1,2]}\cdot e^{\frac{5}{2}\lambda^2}, \nn \\
\sigma_{[1,1,2]}-3\sigma_{[1,1,1,1]} = P_{[1,1,2]}\cdot e^{3\lambda^2}, \nn \\
\sigma_{[1^{m-2},2]}-(m-1)\sigma_{[1^m]} = P_{[1^{m-2},2]}\cdot e^{\frac{m+2}{2}\lambda^2}
\ee
Thus
\be
\sigma_{[1,1 ]} = P_{[1,1 ]}\cdot e^{ \lambda^2}, \nn \\
\sigma_{[ 2]}- \sigma_{[1,1 ]} = P_{[2]}\cdot e^{2\lambda^2}
\label{sigmavsP2}
\ee
\be
\sigma_{[1,1,1]} = P_{[1,1,1]}\cdot e^{\frac{3}{2}\lambda^2}, \nn \\
\sigma_{[1,2]}-2\sigma_{[1,1,1]} = P_{[1,2]}\cdot e^{\frac{5}{2}\lambda^2}, \nn \\
\sigma_{[3]}-\sigma_{[1, 2]}+\sigma_{[1,1,1]} = P_{[3]}\cdot e^{\frac{9}{2}\lambda^2}
\ee
\be
\sigma_{[1,1,1,1]} = P_{[1,1,1,1]}\cdot e^{2\lambda^2}, \nn \\
\sigma_{[1,1,2]}-3\sigma_{[1,1,1,1]} = P_{[1,1,2]}\cdot e^{3\lambda^2}, \nn \\
\sigma_{[2,2]}-\sigma_{[1,1,2]}+\sigma_{[1,1,1,1]} = P_{[2,2]}\cdot e^{4\lambda^2}, \nn \\
\sigma_{[1,3]} -\sigma_{[2,2]}-\sigma_{[1,1,2]} + 2\sigma_{[1,1,1,1]} = P_{[1,3]}\cdot e^{5\lambda^2}, \nn \\
\sigma_{[4]}-\sigma_{[1,3]} +\sigma_{[1,1,2]} -\sigma_{[1,1,1,1]} = P_{[4]}\cdot e^{8\lambda^2}
\ee

\be
\sigma_{[1,1,1,1,1]} = P_{[1,1,1,1,1]}\cdot e^{\frac{5}{2}\lambda^2}, \nn \\
\sigma_{[1,1,1,2]}-4\sigma_{[1,1,1,1,1]} = P_{[1,1,1,2]}\cdot e^{\frac{7}{2}\lambda^2}, \nn \\
\sigma_{[1,2,2]}-2\sigma_{[1,1,1,2]}+3\sigma_{[1,1,1,1,1]} = P_{[1,2,2]}\cdot e^{\frac{9}{2}\lambda^2}, \nn \\
\sigma_{[1,1,3]} - \sigma_{[1,2,2]}-\sigma_{[1,1,1,2]}+3\sigma_{[1,1,1,1,1]} = P_{[1,1,3]}\cdot e^{\frac{11}{2}\lambda^2}, \nn \\
\sigma_{[2,3]}-\sigma_{[1,1,3]} - \sigma_{[1,2,2]}+2\sigma_{[1,1,1,2]}-2\sigma_{[1,1,1,1,1]} = P_{[2,3]}\cdot e^{\frac{13}{2}\lambda^2}, \nn \\
\sigma_{[1,4]}-\sigma_{[2,3]}-\sigma_{[1,1,3]} + \sigma_{[1,2,2]}+\sigma_{[1,1,1,2]}-2\sigma_{[1,1,1,1,1]} = P_{[1,4]}\cdot e^{\frac{17}{2}\lambda^2}, \nn \\
\sigma_{[5]}-\sigma_{[1,4]}+\sigma_{[1,1,3]} -\sigma_{[1,1,1,2]}+\sigma_{[1,1,1,1,1]} = P_{[5]}\cdot e^{\frac{25}{2}\lambda^2}
\ee
At the next level we encounter a new phenomenon -- degeneracy of eigenvalues $\mu$,
therefore the two lines with bullets are ambiguous, one can add the preceding line with any coefficient.
We make the choice. consistent with the formulas (\ref{sigvsA6}) below.
With this choice the coefficient in the box vanishes -- thus passing our test for ordering independence
(see below).
\be
\sigma_{[1^6]} = P_{[1^6]}\cdot e^{3\lambda^2},
\nn \\
\sigma_{[1^4,2]}-5 \sigma_{[1^6]} = P_{[1^4,2]}\cdot e^{4\lambda^2},
\nn \\
\sigma_{[1,1,2,2]}-3\sigma_{[1^4,2]}+6 \sigma_{[1^6]} = P_{[1,1,2,2]}\cdot e^{5\lambda^2},
\nn \\
\sigma_{[2,2,2]}- \sigma_{[1,1,2,2]}+\sigma_{[1^4,2]}-  \sigma_{[1^6]}
= P_{[2,2,2]}\cdot e^{6\lambda^2},
\nn \\
\bullet \ \ \sigma_{[1,1,1,3]}-\boxed{0}\cdot \sigma_{[2,2,2]}- \sigma_{[1,1,2,2]}- \sigma_{[1^4,2]}+4 \sigma_{[1^6]}
= P_{[1,1,3]}\cdot e^{6\lambda^2},
\nn \\
\sigma_{[1,2,3]}-2\sigma_{[1,1,1,3]}-2\sigma_{[2,2,2]}+ 0\cdot \sigma_{[1,1,2,2]}+4\sigma_{[1^4,2]}-6\sigma_{[1^6]}
= P_{[1,2,3]}\cdot e^{7\lambda^2},
\nn \\
\sigma_{[3,3]}-\sigma_{[1,2,3]}+ \sigma_{[1,1,1,3]}+  \sigma_{[2,2,2]}+0\cdot\sigma_{[1,1,2,2]}
- \sigma_{[1^4,2]}+  \sigma_{[1^6]}
= P_{[3,3]}\cdot e^{9\lambda^2},
\nn \\
\bullet \ \  \sigma_{[1,1,4]}+0\cdot \sigma_{[3,3]}- \sigma_{[1,2,3]}- \sigma_{[1,1,1,3]}+  \sigma_{[2,2,2]}+\sigma_{[1,1,2,2]}
+ \sigma_{[1^4,2]}-3 \sigma_{[1^6]}
= P_{[1,1,4]}\cdot e^{9\lambda^2},
\nn \\
\sigma_{[2,4]}-\sigma_{[1,1,4]}-\sigma_{[3,3]}+0\cdot\sigma_{[1,2,3]}+ \sigma_{[1,1,1,3]} -\sigma_{[2,2,2]}+\sigma_{[1,1,2,2]}-2\sigma_{[1^4,2]}+2 \sigma_{[1^6]}
= P_{[2,4]}\cdot e^{10\lambda^2},
\nn \\
\sigma_{[1,5]}-\sigma_{[2,4]}- \sigma_{[1,1,4]}+0\cdot \sigma_{[3,3]}+ \sigma_{[1,2,3]}+ \sigma_{[1,1,1,3]}
+ 0\cdot \sigma_{[2,2,2]}-  \sigma_{[1,1,2,2]}- \sigma_{[1^4,2]}+2 \sigma_{[1^6]}
= P_{[1,5]}\cdot e^{13\lambda^2},
\nn
\ee
\vspace{-0.7cm}
\be
\sigma_{[6]}-\sigma_{[1,5]}+\sigma_{[1,1,4]}-\sigma_{[1,1,1,3]} +\sigma_{[1,1,1,1,2]}-\sigma_{[1,1,1,1,1,1]}
= P_{[6]}\cdot e^{18\lambda^2}
\label{sigmavsP6}
\ee
Therefore
\be
\sigma_{[1,1,1]} = P_{[1,1,1]}\cdot e^{\frac{3}{2}\lambda^2}, \nn \\
\sigma_{[1,2]} =P_{[1,2]}\cdot e^{\frac{5}{2}\lambda^2} + 2\sigma_{[1,1,1]} =
P_{[1,2]}\cdot e^{\frac{5}{2}\lambda^2} + 2P_{[1,1,1]}\cdot e^{\frac{3}{2}\lambda^2}, \nn \\
\sigma_{[3]} = L_{N-1}^1(-m^2\lambda^2)e^{\frac{9}{2}\lambda^2} + \sigma_{[1,2]} - \sigma_{[1,1,1]}
= L_{N-1}^1(-m^2\lambda^2)e^{\frac{9}{2}\lambda^2}
+P_{[1,2]}\cdot e^{\frac{5}{2}\lambda^2} + P_{[1,1,1]}\cdot e^{\frac{3}{2}\lambda^2}
\label{sigmaviaP3}
\ee

\be
\sigma_{[1,1,1,1]} =  P_{[1,1,1,1]}\cdot e^{2\lambda^2} \nn \\
\sigma_{[1,1,2]} = P_{[1,1,2]}\cdot e^{3\lambda^2} +3 P_{[1,1,1,1]}\cdot e^{2\lambda^2} \nn \\
\sigma_{[2,2]} =  P_{[2,2]}\cdot e^{4\lambda^2}+ P_{[1,1,2]}\cdot e^{3\lambda^2}
+2 P_{[1,1,1,1]}\cdot e^{2\lambda^2} \nn \\
\sigma_{[1,3]} = P_{[1,3]}\cdot e^{5\lambda^2}+P_{[2,2]}\cdot e^{4\lambda^2}+ 2P_{[1,1,2]}\cdot e^{3\lambda^2}
+3 P_{[1,1,1,1]}\cdot e^{2\lambda^2} \nn \\
\sigma_{[4]} = P_{[4]}\cdot e^{8\lambda^2} + P_{[1,3]}\cdot e^{5\lambda^2}+P_{[2,2]}\cdot e^{4\lambda^2}
+ P_{[1,1,2]}\cdot e^{3\lambda^2} + P_{[1,1,1,1]}\cdot e^{2\lambda^2}
\ee

\be
\sigma_{[1,1,1,1,1]} =  P_{[1,1,1,1,1]}\cdot e^{\frac{5}{2}\lambda^2}
\nn \\
\sigma_{[1,1,1,2]} = P_{[1,1,1,2]}\cdot e^{\frac{7}{2}\lambda^2} +4  P_{[1,1,1,1,1]}\cdot e^{\frac{5}{2}\lambda^2}
\nn \\
\sigma_{[1,2,2]} =  P_{[1,2,2]}\cdot e^{\frac{9}{2}\lambda^2}+ 2P_{[1,1,1,2]}\cdot e^{\frac{7}{2}\lambda^2}
+5  P_{[1,1,1,1,1]}\cdot e^{\frac{5}{2}\lambda^2}
\nn \\
\sigma_{[1,1,3]} = P_{[1,1,3]}\cdot e^{\frac{11}{2}\lambda^2}+P_{[1,2,2]}\cdot e^{\frac{9}{2}\lambda^2}
+ 3P_{[1,1,1,2]}\cdot e^{\frac{7}{2}\lambda^2}+6 P_{[1,1,1,1,1]}\cdot e^{\frac{5}{2}\lambda^2}
\nn \\
\sigma_{[2,3]} = P_{[2,3]}\cdot e^{\frac{13}{2}\lambda^2}
+ P_{[1,1,3]}\cdot e^{\frac{11}{2}\lambda^2}+2P_{[1,2,2]}\cdot e^{\frac{9}{2}\lambda^2}
+ 3P_{[1,1,1,2]}\cdot e^{\frac{7}{2}\lambda^2} + 5 P_{[1,1,1,1,1]}\cdot e^{\frac{5}{2}\lambda^2}
\nn \\
\sigma_{[1,4]} = P_{[1,4]}\cdot e^{\frac{17}{2}\lambda^2}  + P_{[2,3]}\cdot e^{\frac{13}{2}\lambda^2}
+2P_{[1,1,3]}\cdot e^{\frac{11}{2}\lambda^2}+ 2P_{[1,2,2]}\cdot e^{\frac{9}{2}\lambda^2}
+ 3P_{[1,1,1,2]}\cdot e^{\frac{7}{2}\lambda^2} + 4 P_{[1,1,1,1,1]}\cdot e^{\frac{5}{2}\lambda^2}
\nn \\
\sigma_{[5]} = P_{[5]}\cdot e^{\frac{25}{2}\lambda^2}\! +P_{[1,4]}\cdot e^{\frac{17}{2}\lambda^2}\!
+ P_{[2,3]}\cdot e^{\frac{13}{2}\lambda^2}\!
+ P_{[1,1,3]}\cdot e^{\frac{11}{2}\lambda^2}\!+ P_{[1,2,2]}\cdot e^{\frac{9}{2}\lambda^2}\!
+ P_{[1,1,1,2]}\cdot e^{\frac{7}{2}\lambda^2}\! +   P_{[1,1,1,1,1]}\cdot e^{\frac{5}{2}\lambda^2}
\nn
\ee
and so on.

\bigskip

In other words, we can represent any $\sigma_R$ as a sum of exponentials with polynomial prefactors
and decomposition is {\it triangular}.
All these prefactors are expressed through Laguerre polynomials, but in a somewhat sophisticated way --
as their multi-linear combinations.
See \cite{MOk} for more details on this interpretation.

\section{Exact statements
\label{exasta}}

The main statement is that there usually is a triangular decomposition
\be
\boxed{
\sigma\!_R = \sum_{Q\leq R} K_{RQ}\, e^{\frac{1}{2}\mu_Q\lambda^2} P_Q(N,\lambda^2)
}
\label{sigmadeco}
\ee
where sum goes over all diagrams $Q$ of the same size as $R$: $|Q|=|R|$.

Now we need to specify four things:
\begin{itemize}
\item{}The {\bf ordering} of Young diagrams $Q$ of a given size.
It is lexicographic, defined by  (2) in \cite{MMkerov}:
\be
\!\!\!\!\!\!\!\!\!\!\!\!\!\!\!\!\!\!\!\!\!\!\!
R>R' \ \ {\rm if} \ \ r_1>r_1' \ \ {\rm or \ \ if} \ \ r_1=r_1', \ \ {\rm but} \ r_2>r_2'
\ \ {\rm or \ if} \ \ r_1=r_1' \ \ {\rm and} \ \ r_2=r_2',\ \ {\rm but} \ \ r_3>r_3' \ \ {\rm and\ so \ on}
\ee
There is ambiguity in this ordering, because
\be
R > R' \ \ {\rm is\ not\ always\ the\ same\ as} \ \  {R'}^\vee > R^\vee
\label{ambig}
\ee
where $R^\vee$ denotes the transposed Young diagram.

\item{} {\bf Triangular matrices $K$} are read from (\ref{sigmavsP2})-(\ref{sigmavsP6}),
converted into the form like (\ref{sigmaviaP3}):
\be
\left(\begin{array}{c} 1 \end{array}\right),
\left(\begin{array}{cc} 1&1 \\&1 \end{array}\right),
\left(\begin{array}{ccc} 1&1&1 \\ &1&2 \\ &&1 \end{array}\right),
\left(\begin{array}{ccccc} 1&1&1&1&1 \\ & 1 &1 &2 &3 \\ &&1 &1 &2 \\ &&&1&3 \\ &&&&1 \end{array}\right),
\left(\begin{array}{ccccccc} 1&1&1&1&1&1&1\\ & 1&1 &2 &2&3&4 \\ && 1 & 1&2&3&5 \\ && &1&1&3&6 \\ &&&&1&2&5  \\ &&&&&1&4 \\ &&&&&&1 \end{array}\right),
\ \ \ldots
\label{sigmaviaPmatrix15}
\ee
The above-mentioned ordering ambiguity \cite{MMkerov} would arise for the first time for $[1,1,1,3]$ and $[2,2,2]$ --
if the corresponding entry
$S_{[1,1,1,3],[2,2,2]}$ and $S_{[2,2,2],[1,1,1,3]}$ were not vanishing.  But they are.
These are marked by boxes in the next matrix in the row:
\be
\!\!\!\!\!
\left(\begin{array}{c } \sigma_{[6]}\\ \sigma_{[1,5]} \\ \sigma_{[2,4]} \\ \sigma_{[1,1,4]} \\ \sigma_{[3,3]}
\\ \sigma_{[1,2,3]} \\ \underline{\sigma_{[1^3,3]}} \\ \underline{\sigma_{[2^3]}}
\\ \sigma_{[1^2,2^2]} \\ \sigma_{[1^4,2]} \\ \sigma_{[1^6]}
\end{array}\right)
= \left(\begin{array}{ccccccccccc} 1&1&1&1&1&1&1&1&1&1&1\\ & 1& 1&2 &1&2&3&2&3&4&5 \\
&& 1 & 1&1&2&3&3&4&6&9 \\ && &1 &0&1&3&1&3&6&10
\\ &&&&1&1&1&1&2&3&5 \\ &&&&&1&2&2&4&8&16 \\ &&&&&&1&\boxed{0}&1&4&10 \\ &&&&&& &1&1&2&5 \\
&&&&&&&&1&3&9 \\ &&&&&&&&&1&5 \\ &&&&&&&&&&1 \end{array}\right)
\Big({\rm diag}(e^{\frac{\mu}{2} \lambda^2})\Big)
\left(\begin{array}{c} P_{[6]} \\P_{[1,5]} \\P_{[2,4]} \\P_{[1,1,4]} \\P_{[3,3]} \\P_{[1,2,3]} \\
\underline{P_{[1,1,1,3]}} \\\underline{P_{[2,2,2]}} \\
P_{[1,1,2,2]} \\ P_{[1,1,1,1,2]}\\ P_{[1,1,1,1,1,1]}
\end{array}\right)
\label{sigmaviaPmatrix6}
\ee
From these examples it is clear that  $K_{RQ}$ coincide with the Kostka matrices \cite{symfun}
for decomposition of Schur symmetric polynomials into the sum
of elementary monomial functions $S_R[X]=\sum_{Q\leq R} K_{RQ}m_Q[X]$ --
which is known to be free of the ordering ambiguity.

\item{} {\bf Eigenvalues} $\mu_Q$, which monotonically interpolate between $\mu_{[m]}=m^2$ and $\mu_{[1^m]}=m$.
Examples of eigenvalue sets for the first six sizes are:
\be
(1), \ (4,2), \ (9,5,3), \ (16,10,8,6,4), \ (25,17,13,11,9,7,5), \ (36,26,20,18,18,14,12,12,10,8,6), \ \ldots
\ \ \
\label{evsets}
\ee
In general, as we will see in sec.\ref{pols} below,
they are sums of squares of integers $\sum_a k_a^2$ for $\sum_a k_a=m$,
ranging from $m$ for $k_1=\ldots=k_m=1$ to $m^2$ for $k_1=m$.

\item{} {\bf Polynomials $P_Q$.}
This is a longer story and we present in  a separate section,
which refers at many places to the examples in the Appendix.
The result is that
the polynomials $P_Q$ are extracted from the expansion (\ref{gensym}) below
by collecting terms with the given exponential prefactors $e^{\frac{\mu_Q\lambda^2}{2}}$.
This understanding will allow to make an even more {\bf concise summary}
in the form of (\ref{summary}) at the very end of the sec.\ref{pols}.

\end{itemize}

\section{Polynomials $P_Q$
\label{pols}}

\begin{itemize}
\item{} First of all, the generating functions are known for polynomials in antisymmetric representations
\cite{1808.10161}:
\be
\sum_{k=0} t^k P_{[1^k]} = {\rm det}_{N\times N}\left(\delta_{i,j} + tL_{i-1}^{j-i}(-\lambda^2) \right)
\ee
where $L_n^a(x) := \frac{e^x}{x^an!} \frac{\p^n (e^{-x}x^{n+a})}{\p x^n}$.
In other words
\be
P_{[1^m]} = S_{[1^m]}\left\{p_k = \tr {\cal L}^k\right\}
\ee
where ${\cal L}_{ij} = L_{i-1}^{j-i}(-\lambda^2)$.
The first answer in (\ref{simplestave}) is reproduced by the amusing identity
for Laguerre polynomials
\be
\sum_{i=1}^N L^0_{i-1} = L^1_{N-1}
\ee

\item{}
Second,  we already know from (\ref{pim}), that
\be
P_{[m]}(N,\lambda^2) =  P_{[1]}(N,m^2\lambda^2) = L_{N-1}^1(-m^2\lambda^2)
\ee
We remind that these are proportional not to averages $\sigma_m$ of symmetric Schurs,
but to those of particular time-variables, $\langle \pi_m \rangle$.

\item{}
Third,
a general formula for symmetric averages is also known \cite{1808.10161}:
\be
\sum_{k=0} t^k \sigma_{[k]} =
{\rm det}_{N\times N} \left(\delta_{i,j}+\sum_{k=1}^\infty t^kA_{ij}(k\lambda)\right)
= {\rm det} \left(1+\sum_k t^k{\cal A}_{k\lambda}\right)
\label{detsym}
\ee
where  {\it symmetric}\footnote{
I am grateful to Kazumi Okuyama for advising symmetric choice of the matrix,
which immediately explains (\ref{tracesym}) below.}
 matrix ${\cal A}_\lambda$ has the entries
\be
A_{ij}(\lambda):= e^{\frac{1}{2}\lambda^2}\sqrt{\frac{(i-1)!}{(j-1)!}}\,\lambda^{j-i}L_{i-1}^{j-i}(- \lambda^2)
\label{Amatrix}
\ee
These are combinations of several exponentials with polynomial coefficients.
In fact, as we already know, all exponentials are contributing to $\sigma_{[m]}$, with unit coefficients:
\be
\boxed{
\sigma_{|m|} = \sum_{Q:\ |Q|=m} e^{\frac{1}{2}\mu_Q\lambda^2} P_Q(N,\lambda^2)
}
\label{symdeco}
\ee
Thus the single formula (\ref{detsym}) is sufficient to extract all $P_Q$ from it.
In particular it is now clear that all the eigenvalue sets in (\ref{evsets})
are sums of squares of integers $\sum_a k_a^2$ for $\sum_a k_a=m$,
ranging from $m$ for $k_1=\ldots=k_m=1$ to $m^2$ for $k_1=m$.

\item{}
The expansion at the r.h.s. of (\ref{detsym}) looks like
\be
{\rm det} \left(1+\sum_k t^k{\cal A}_{k\lambda}\right) =
1 + t\cdot \overbrace{\Tr {\cal A}_\lambda}^{\mathfrak{A}_{[1]}}
+ t^2\Big\{\overbrace{\Tr {\cal A}_{2\lambda}}^{\mathfrak{A}_{[2]}} +
\overbrace{\frac{(\Tr {\cal A}_\lambda)^2-\Tr {\cal A}_\lambda^2}{2}}^{\mathfrak{A}_{[1,1]}}\Big\}
+ \nn \\ \!\!\!\!
+ t^3\Big\{\underbrace{\Tr {\cal A}_{3\lambda}}_{\mathfrak{A}_{[3]}} +
\underbrace{\Big(\Tr {\cal A}_{2\lambda} \,\Tr{\cal A}_\lambda
- \Tr({\cal A}_{2\lambda}{\cal A}_\lambda)\Big)}_{\mathfrak{A}_{[1,2]}}+
\underbrace{\frac{(\Tr {\cal A}_\lambda)^3-3\Tr {\cal A}_{\lambda}^2\,\Tr{\cal A}_\lambda
+\Tr {\cal A}_\lambda^3}{6}}_{\mathfrak{A}_{[1,1,1]}}
\Big\}
+ \ldots
= \sum_Q t^{|Q|}\mathfrak{A}_Q
\label{gensym}
\ee
Combinations $\mathfrak{A}_Q$ here are the ones, which have the same exponential prefactors.
They are labeled by partitions $Q=\left\{0<k_1\leq \ldots \leq k_{l_Q}\right\}$
and consist of various traces of matrices ${\cal A}_{k_a\lambda}$ with the same value of
$\mu_Q = \sum_a k_a^2$.
These $\mathfrak{A}_Q$ are essentially the polynomials of our interest:
$\mathfrak{A}_Q = e^{\frac{\mu_Q\lambda^2}{2}}P_Q$.
In fact we can use them to make  (\ref{sigmadeco}) even shorter:
\be
\boxed{\boxed{
\sigma\!_R = \sum_{Q\leq R} K_{RQ}\,\mathfrak{A}_Q
}}
\label{sigmadecoU}
\ee
Despite in the first examples above $\mathfrak{A}_Q$ can seem similar to Schur polynomials,
this is true only for $\mathfrak{A}_{[m]}$ and $\mathfrak{A}_{[1^m]}$, the rest contain products
of different matrices under the trace signs, and can not be expressed through
variables like $\Tr {\cal A}_{k\lambda}$ -- as illustrated already by the example of $\mathfrak{A}_{[1,2]}$.

\bigskip

{\bf Examples:}

\item{}
In this way we get what we already know from the Appendix:

{\bf Levels 1-3:}  \vspace{-0.5cm}
\be
P_{[1]} = e^{-\frac{\lambda^2}{2}}\cdot \Tr {\cal A}_\lambda
\nn \\ \nn \\
P_{[2]} = e^{-2\lambda^2}\cdot \Tr {\cal A}_{2\lambda}  \nn \\
P_{[1,1]} = e^{-\lambda^2}\cdot \frac{(\Tr{\cal A}_\lambda)^2-\Tr {\cal A}_\lambda^2}{2}
\nn \\ \nn \\
P_{[3]} = e^{-\frac{9\lambda^2}{2}}\cdot \Tr {\cal A}_{3\lambda}  \nn \\
P{[1,1,1]} = e^{-\frac{3\lambda^2}{2}}\cdot
\frac{(\Tr {\cal A}_\lambda)^3 - 3\,\Tr {\cal A}_\lambda^2\, \Tr{\cal A}_\lambda
+ 2\,\Tr {\cal A}_\lambda^3}{6}
\label{sigvsA13}
\ee
Note that ${\cal A}_\lambda = {\cal L}$
and single traces $\Tr {\cal A}_{k\lambda} = \Tr {\cal L}\Big|_{\lambda=k\lambda}$,
thus formulas for $P_{[1^m]}$ and $P_{[m]}$ do not change.
But at level three we get also a new formula:
\be
P_{[1,2]} = e^{-\frac{5\lambda^2}{2}}\cdot \Big(\Tr {\cal A}_{2\lambda} \,\Tr {\cal A}_{\lambda} -
\Tr ({\cal A}_{2\lambda}{\cal A}_{\lambda})\Big)
\ee

\bigskip

At higher levels the  expressions for the polynomials
$P_Q = e^{-\frac{\mu_Q\lambda^2}{2}}\mathfrak{A}_Q$ are:

{\bf Level\ 4:}
\be
P_{[4]} = e^{-8\lambda^2}\cdot \Tr {\cal A}_{4\lambda} \nn \\
P_{[1,3]} = e^{-5\lambda^2}\cdot \Big(\Tr {\cal A}_{3\lambda} \,\Tr {\cal A}_{\lambda} -
\Tr ({\cal A}_{3\lambda}{\cal A}_{\lambda})\Big)  \nn \\
P_{[2,2]} =  e^{-4\lambda^2}\cdot \left(\frac{(\Tr {\cal A}_{2\lambda})^2   -
\Tr ({\cal A}_{2\lambda}^2)}{2}\right)  \nn \\
P_{[1,1,2]} =  e^{-3\lambda^2}\cdot
\left(\frac{(\Tr {\cal A}_\lambda)^2-\Tr{\cal A}_\lambda^2}{2}\, \Tr {\cal A}_{2\lambda}
-\Tr ({\cal A}_{2\lambda}{\cal A}_\lambda)\,\Tr {\cal A}_\lambda
+ \Tr({\cal A}_{2\lambda}{\cal A}_\lambda^2)\right) \nn \\
P_{[1,1,1,1]} = e^{-2\lambda^2} \cdot
S_{[1,1,1,1]}\{p_k= \Tr{\cal A}_\lambda^k\}
\ee

\bigskip

{\bf Level\ 5:}
\be
P_{[5]} = e^{-\frac{25}{2}\lambda^2}\cdot \Tr {\cal A}_{5\lambda}
\nn \\
P_{[1,4]} = e^{-\frac{17}{2}\lambda^2}\cdot \Big(\Tr {\cal A}_{4\lambda} \,\Tr {\cal A}_{\lambda} -
\Tr ({\cal A}_{4\lambda}{\cal A}_{\lambda})\Big)
\nn \\
P_{[2,3]} =  e^{-\frac{13}{2}\lambda^2}\cdot \Big(\Tr {\cal A}_{3\lambda} \,\Tr {\cal A}_{2\lambda} -
\Tr ({\cal A}_{3\lambda}{\cal A}_{2\lambda})\Big)
\nn \\
P_{[1,1,3]} =  e^{-\frac{11}{2}\lambda^2}\cdot
\left(\frac{(\Tr {\cal A}_\lambda)^2-\Tr{\cal A}_\lambda^2}{2}\, \Tr {\cal A}_{3\lambda}
-\Tr ({\cal A}_{3\lambda}{\cal A}_\lambda)\,\Tr {\cal A}_\lambda
+ \Tr({\cal A}_{3\lambda}{\cal A}_\lambda^2)\right)
\nn \\
P_{[1,2,2]} =  e^{-\frac{9}{2}\lambda^2}\cdot
\left(\frac{(\Tr {\cal A}_{2\lambda})^2-\Tr{\cal A}_{2\lambda}^2}{2}\, \Tr {\cal A}_{\lambda}
-\Tr ({\cal A}_{2\lambda}{\cal A}_\lambda)\,\Tr {\cal A}_{2\lambda}
+ \Tr({\cal A}_{\lambda}{\cal A}_{2\lambda}^2)\right)
\nn \\
\!\!\!\!\!\!\!\!\!\!\!\!\!\!\!\!\!\!\!\!
P_{[1,1,1,2]} = e^{-\frac{7}{2}\lambda^2}\cdot
\Big(
\frac{ (\Tr{\cal A}_\lambda)^3
-3 \Tr{\cal A}_\lambda^2\,\Tr{\cal A}_\lambda + 2\Tr{\cal A}_\lambda^3}{6}\,\Tr{\cal A}_{2\lambda}
-\frac{(\Tr {\cal A}_{\lambda})^2-\Tr{\cal A}_{\lambda}^2}{2}\, \Tr ({\cal A}_{2\lambda}{\cal A}_{\lambda})
-\nn \\
-\Tr ({\cal A}_{2\lambda}{\cal A}_\lambda^3)
+ \Tr({\cal A}_{2\lambda}{\cal A}_{\lambda}^2)\,\Tr{\cal A}_\lambda
\Big)
\nn \\
P_{[1,1,1,1,1]} = e^{-\frac{5}{2}\lambda^2} \cdot S_{[1,1,1,1,1]}\{p_k= \Tr{\cal A}_\lambda^k\}
\ee

\bigskip

At the next {\bf level 6} we encounter a number of new phenomena, partly mentioned in the text above:
\begin{itemize}
\item{}
There is a degeneracy of eigenvalues $\mu$, what makes decomposition of $\sigma$ ambiguous --
but expression through matrices ${\cal A}$ fixes the ambiguity.
\item{}
This fixation provides the answer, which is free from the ordering ambiguity from \cite{MMkerov},
at least at the level 6, where it could appear for the first time --
the coefficients in the boxes in (\ref{sigmavsP6}) and (\ref{sigmaviaPmatrix6}) are zero.
\item{}
The matrices ${\cal A}_{k\lambda}$ with different $k$ do not commute,
what makes essential their order within traces.

Again this happens for the first time at level 6 -- in two places.
The first -- double-underlined pair in (\ref{sigvsA6}) below,--
appears to be a false alarm: in fact,
due to symmetricity of the matrix (\ref{Amatrix})
\be
\Tr {\cal A}_{\lambda} {\cal A}_{2\lambda}{\cal A}_{3\lambda} \ \stackrel{{\cal A}^{\rm tr}={\cal A}}{=}\
\Tr {\cal A}_{3\lambda}{\cal A}_{2\lambda}{\cal A}_{\lambda}
\ = \ \Tr {\cal A}_{\lambda}{\cal A}_{3\lambda}{\cal A}_{2\lambda}
\label{tracesym}
\ee
However,  in the next --  triple-underlined -- example the ordering matters even in traces:
$$\Tr {\cal A}_{\lambda}{\cal A}_{2\lambda}{\cal A}_{\lambda}{\cal A}_{2\lambda}
\neq \Tr {\cal A}_{\lambda}^2{\cal A}_{2\lambda}^2$$

\end{itemize}

\newpage
{\bf Level\ 6:}
\be
P_{[6]} = e^{-18\lambda^2}\cdot \Tr {\cal A}_{6\lambda}
\nn \\ \nn \\
P_{[1,5]} = e^{-13\lambda^2}\cdot \Big(\Tr {\cal A}_{5\lambda} \,\Tr {\cal A}_{\lambda} -
\Tr ({\cal A}_{5\lambda}{\cal A}_{\lambda})\Big)
\nn \\ \nn \\
P_{[2,4]} =  e^{-10\lambda^2}\cdot \left(\Tr {\cal A}_{4\lambda}\, \Tr {\cal A}_{2\lambda}   -
\Tr ({\cal A}_{4\lambda}{\cal A}_{2\lambda})\right)
\nn \\ \nn \\
P_{[1,1,4]} =  e^{-9\lambda^2}\cdot
\left(\frac{(\Tr {\cal A}_\lambda)^2-\Tr{\cal A}_\lambda^2}{2}\, \Tr {\cal A}_{4\lambda}
-\Tr ({\cal A}_{4\lambda}{\cal A}_\lambda)\,\Tr {\cal A}_\lambda
+ \Tr({\cal A}_{4\lambda}{\cal A}_\lambda^2)\right)
\nn\\ \nn \\
P_{[3,3]} =  e^{-9\lambda^2}\cdot \left(\frac{(\Tr {\cal A}_{3\lambda})^2   -
\Tr ({\cal A}_{3\lambda}^2)}{2}\right)
\nn \\ \nn \\
P_{[1,2,3]} =  e^{-7\lambda^2}\cdot \Big(
\underline{\underline{
 \Tr ({\cal A}_{\lambda}{\cal A}_{2\lambda}{\cal A}_{3\lambda}) +
\Tr ({\cal A}_{\lambda}{\cal A}_{3\lambda}{\cal A}_{2\lambda})
}} - \nn \\
-\Tr ({\cal A}_{3\lambda}{\cal A}_{2\lambda})\,\Tr{\cal A}_\lambda
-\Tr ({\cal A}_{3\lambda}{\cal A}_{1\lambda})\,\Tr{\cal A}_{2\lambda}
-\Tr ({\cal A}_{2\lambda}{\cal A}_{1\lambda})\,\Tr{\cal A}_{3\lambda}
+ \Tr {\cal A}_{1\lambda}\,\Tr{\cal A}_{2\lambda}\,\Tr{\cal A}_{3\lambda}
\Big)
\nn \\ \nn \\
\!\!\!\!\!\!\!\!\!\!\!
P_{[1,1,1,3]} =  e^{-6\lambda^2}\cdot \Big(
\frac{ (\Tr{\cal A}_\lambda)^3
-3 \Tr{\cal A}_\lambda^2\,\Tr{\cal A}_\lambda + 2\Tr{\cal A}_\lambda^3}{6}\,\Tr{\cal A}_{3\lambda}+
\nn \\ +
\frac{\Tr {\cal A}_{3\lambda} \Tr {\cal A}_{\lambda}-\Tr({\cal A}_{3\lambda}{\cal A}_\lambda)}{2}
\, \Tr ( {\cal A}_{\lambda}^2)
- \Tr ({\cal A}_{3\lambda}{\cal A}_\lambda^3) \Big)
\nn \\
P_{[2,2,2]} =  e^{-6\lambda^2}\cdot \left(
\frac{ (\Tr{\cal A}_{2\lambda})^3
-3 \Tr{\cal A}_{2\lambda}^2\,\Tr{\cal A}_{2\lambda} + 2\Tr{\cal A}_{2\lambda}^3}{6}\right)
\nn \\ \nn \\
P_{[1,1,2,2]} =  e^{-5\lambda^2}\cdot \left(
\Tr ({\cal A}_{2\lambda}^2{\cal A}_\lambda)\,\Tr{\cal A}_\lambda
+\Tr ({\cal A}_{2\lambda}{\cal A}_\lambda^2)\,\Tr{\cal A}_{2\lambda}
+ \frac{(\Tr {\cal A}_{2\lambda}{\cal A}_\lambda)^2}{2}
+\right. \nn \\ \left.   \!\!\!\!\!\!\!\!\!\!\!\!\!\!\!\!\!\!\!\!\!\!
+ \frac{(\Tr{\cal A}_{2\lambda})^2-\Tr{\cal A}_{2\lambda}^2}{2}
\cdot\frac{(\Tr{\cal A}_{\lambda})^2-\Tr{\cal A}_\lambda^2}{2}
\,\underline{\underline{\underline{
- \Tr({\cal A}_{2\lambda}^2{\cal A}_\lambda^2)
- \frac{1}{2}\Tr ({\cal A}_{2\lambda}{\cal A}_\lambda{\cal A}_{2\lambda}{\cal A}_\lambda)
}}}
- \Tr({\cal A}_{2\lambda}{\cal A}_\lambda)\,\Tr {\cal A}_{2\lambda}\,\Tr {\cal A}_\lambda
\right)
\nn \\ \nn \\
P_{[1,1,1,1,2]} =  e^{-4\lambda^2}\cdot \left(
\frac{(\Tr{\cal A}_\lambda)^4 -6\Tr {\cal A}_{\lambda}^2\,(\Tr {\cal A}_\lambda)^2  + 3(\Tr{\cal A}_\lambda^2)^2
+8\Tr {\cal A}_\lambda^3\,\Tr{\cal A}_\lambda - 6\Tr {\cal A}_\lambda^4}{24}\,\Tr {\cal A}_{2\lambda}
- \right. \nn \\ \left.
- \frac{(\Tr {\cal A}_\lambda)^3-3\Tr{\cal A}_\lambda^2\,\Tr{\cal A}_\lambda
+ 2\Tr {\cal A}_\lambda^3}{6}\,\Tr({\cal A}_{2\lambda}{\cal A}_\lambda)
+\frac{(\Tr{\cal A}_{\lambda})^2-\Tr{\cal A}_\lambda^2}{2}\,\Tr({\cal A}_{2\lambda}{\cal A}_\lambda^2)
- \right.\nn \\ \left.
- \Tr{\cal A}_\lambda\,\Tr({\cal A}_{2\lambda}{\cal A}_\lambda^3)
+ \Tr ({\cal A}_{2\lambda}{\cal A}_\lambda^4)
\right)
\nn \\ \nn \\
P_{[1^6]} = e^{-3\lambda^2} \cdot S_{[1^6]}\{p_k= \Tr{\cal A}_\lambda^k\} \ \ \ \
\label{sigvsA6}
\ee

\end{itemize}

\bigskip

{\bf Summary:}

\bigskip

Formulas (\ref{sigvsA13})-(\ref{sigvsA6}) provide a precise reformulation of mnemonic relations
from sec.\ref{exa}, and seem sufficient to illustrate the general prescription:

\bigskip

\be
\boxed{\boxed{
\text{\bf Extract $\mathfrak{A}_Q$ from (\ref{gensym}) and insert them into (\ref{sigmadecoU})
with the Kostka matrix $K$}
}}
\label{summary}
\ee

\newpage

\section{Conclusion
\label{conc}}

In this paper we described a way to calculate Gaussian averages $\sigma_R$ of exponential functions
$S_R[\tr e^{\lambda X}]$.
Despite Gaussian integral of a linear exponential looks trivial,
in the case of matrices it is not so, and this calculation is a rather involved and even tedious task.
We do it with the help of superintegrability property \cite{SI}, which provides exact
expression for the averages $S_R[X]$ which strongly enhances integrability
in the theory of Gaussian matrix models \cite{UFN23,Mir}
and is responsible for the distinguished role of
Schur polynomials $S_R\{p_k\}$ in them.

Our main result (\ref{sigmadeco}) is that the average $\sigma_R$ is a triangular sum over partitions
$Q\leq R$, each contributing a simple exponential factor $e^{\frac{1}{2}\mu_Q\lambda^2}$
and a sophisticated polynomial prefactor $P_Q(\lambda^2)$.
The previous results \cite{0010274}, \cite{1808.10161} calculated symmetric and antisymmetric averages
$\sigma_{[m]}$ and $\sigma_{[1^m]}$
with the help of ancient orthogonal polynomials method,
and expressed these quantities through Laguerre polynomials.
Decomposition (\ref{sigmadeco}) with the property (\ref{symdeco})
actually allows to extract arbitrary $\sigma_Q$ from $\sigma_{[m]}$
and thus it is also expressed through Laguerres.

This expression, however, is {\it not} simple for a variety of reasons.
The main drawback is the lack of explicit analytical dependence on $N$ --
it enters trough the size of traces and is not easy to extract.
Also, because the answers depend on traces of a few different matrices
they are {\it not} reduced to Schur-like polynomials of a single set of time variables.
Moreover, matrices do not commute and the set of independent traces is rather huge,
{\it not} labeled even by partitions, i.e. the {\it ordered} sets of integers $k_a$,
like in \cite{MTs}.
Instead the would-be time-variables are labeled by {\it non-ordered} sequences,
i.e. by {\it weak compositions}, which appeared in the recent series of papers \cite{MMP} --
though not yet in the role of time labels.
Thus there is still room for a search for better formulation and closed formulas.
An additional ``small'' question here concerns the ordering in triangular decomposition \cite{MMkerov} --
at the first level 6, where ambiguity could show up, it turns to be absent,
but it is not fully clear if this persists at all higher levels.
In other words, the interpretation of $K_{RQ}$ in (\ref{sigmadeco})
as Kostka matrices remains to be explained and justified.
There is still what to do along the lines of this paper.

\section*{Acknowledgements}

I am indebted to V.Mishnyakov for advertising to me the problem of exponential
Gaussian averages and for pointing the references \cite{0010274} and \cite{1808.10161}.
This work is supported by the RSF grant 24-12-00178.

\end{document}